\documentclass[prl, twocolumn, showpacs, superscriptaddress]{revtex4-1}
\usepackage{graphicx,amsmath}

\begin{document}

\title{Realizing the Haldane Phase with Bosons in Optical Lattices}
\author{Junjun Xu}
\affiliation{Department of Physics and Beijing Key Laboratory for Magneto-Photoelectrical Composite and Interface Science, University of Science and Technology Beijing, Beijing 100083, China}
\affiliation{Laboratory of Atomic and Solid State Physics, Cornell University, Ithaca, New York 14853, USA}
\author{Qiang Gu}\email{qgu@ustb.edu.cn}
\affiliation{Department of Physics and Beijing Key Laboratory for Magneto-Photoelectrical Composite and Interface Science, University of Science and Technology Beijing, Beijing 100083, China}
\author{Erich J. Mueller}\email{em256@cornell.edu}
\affiliation{Laboratory of Atomic and Solid State Physics, Cornell University, Ithaca, New York 14853, USA}
\date{\today}

\begin{abstract}
We analyze an experimentally realizable model of bosons in a zig-zag optical lattice, showing that by rapidly modulating the magnetic field one can tune interaction parameters and realize an analog of the Haldane phase. We explain how quantum gas microscopy can be used to detect this phase's non-local string order and its topological edge states.  We model the detection process. We also find that this model can display supersolid correlations, but argue that they only occur at parameter values which would be challenging to realize in an experiment.
\end{abstract}

\maketitle

In the past 30 years, one of the dominant themes in condensed matter theory has been the search for models where the collective excitations behave unlike any known fundamental particle.  While many such fractionalized and topologically ordered models have been found \cite{topology}, very few of them have been experimentally realized.  Here we show how to build on a  setup proposed by the NIST cold atom experimental group \cite{Spielman} to explore one of the iconic fractionalized phases, the Haldane phase of a spin-1 chain \cite{Haldane}.

In 1983, Haldane showed that the properties of integer and half-integer spin chains can be profoundly different \cite{Haldane, review}.  Over the following decade, several researchers explored the rich properties of the integer spin chain, finding half integer spin edge modes \cite{AKLT, Hagiwara, Glarum}, and non-local string order \cite{Nijs, Girvin, Kennedy}.  More recently, Dalla Torre, Berg, and Altman noted that similar physics should occur for spinless bosons hopping on a one-dimensional lattice:  the occupation numbers on each site plays the role of the different spin states \cite{Altman, Altman2}. Subsequently, analogs of the Haldane phase have been predicted for a number of one-dimensional Bose-Hubbard models with off-site interactions \cite{Rossini, Batrouni, Ejima, Lange}.  One enlarges the parameter range over which the Haldane phase is stable if there is a constraint on the maximum number of particles per site.  By combining a number of experimental techniques, we show how to realize a model which would be expected to support the Haldane phase.
We use Density Matrix Renormalization Group (DMRG) techniques to calculate the properties of this model \cite{White, Schollwock}, and explain how to detect the exotic signatures of the Haldane phase.

In a system of one-dimensional lattice bosons, the Haldane (HI) phase lies at the intersection of the density wave (DW) phase, where double occupied sites (doublons) alternate with empty sites (holons), the Mott insulator (MI) phase, where each site is occupied by a single atom, and the superfluid (SF) phase, where the quasiparticles (doublons and holons) are free to move around.  In the Haldane phase the quasiparticles are fluid but ordered: their spacing varies, but as one moves from left to right the next quasiparticle after a doublon is a holon, and vice-versa.  This ground state is four-fold degenerate in a large but finite system with hard-wall boundary conditions -- corresponding to the flavors of the leftmost and rightmost quasiparticles -- which are bound to the edges of the system.  This four-fold degeneracy was also found in the original spin context, corresponding to two spin-1/2 degrees of freedom, one sitting at each boundary.

\begin{figure}[t]
  \includegraphics[width=0.45\textwidth]{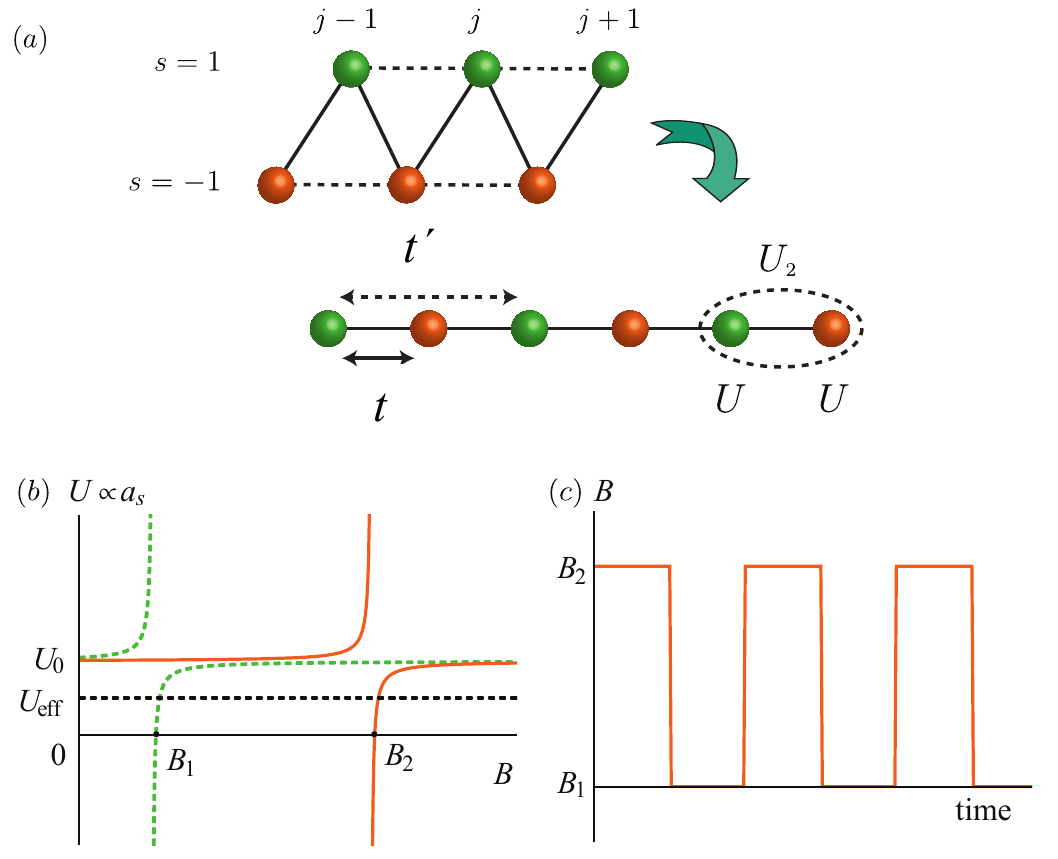}
  \caption{(Color Online)
  (a) Schematic of the experimental system, which can be interpreted as a zig-zag ladder or a 1D lattice with next-nearest neighbor hopping. The green and orange colors label two different spin states $s=\pm 1$.  (b) On-site interaction $U$ in $F=1$, $m_F=0$ (green dashed) and $m_F=1$ (solid orange) hyperfine states of $^{87}$Rb.  The background interaction strength $U_0$ corresponds to the value of $U$ away from the Feshbach resonances near the zero-crossings at $B_1$ and $B_2$.  Rapidly switching the magnetic field between $B_1$ and $B_2$, as illustrated in (c), yields an effective time-averaged on-site interaction $U_{\rm eff}=U_0/2$ in both channels.}
  \label{fig:fig1}
\end{figure}

One-dimensional bosonic system have been realized by trapping cold atoms in elongated optical traps \cite{exp1,exp2,exp3}. Anisimovas {\it et al}. showed that by using a one-dimensional (1D) spin dependent optical lattice and Raman induced hopping, one could produce the zig-zag lattice illustrated in Fig. \ref{fig:fig1}(a), described by the tight-binding model \cite{Spielman}
\begin{align}\label{eq:zigzag}
H&=t\sum_j(c^\dagger_{1,j}c_{-1,j}+c^\dagger_{1,j-1}c_{-1,j}+{\rm H.c.})\nonumber\\
&-t^\prime\sum_{j,s}(c^\dagger_{s,j+1}c_{s,j}+c^\dagger_{s,j}c_{s,j+1})+\frac{U}{2}\sum_{j,s}n_{s,j}(n_{s,j}-1)\nonumber\\
&+U_2\sum_j[n_{1,j}+n_{1,j-1}]n_{-1,j}.
\end{align}
Here $s=\pm1$ labels the spin state of the atoms and $j$ is the position of the atom along the lattices.
The hopping between and within these two spin states are characterized by $t$ and $t^\prime$. The on-site and nearest interspecies interaction are described by $U$ and $U_2$.
Following Anisimovas {\it et al}. and shown in Fig. \ref{fig:fig1}(a), one can think of Eq.~(\ref{eq:zigzag}) in one of two ways: either as a two-leg ``zig-zag" lattice, or a 1D chain with next-nearest neighbor hopping.   The latter is closer to the actual physical system.  Following that interpretation, we introduce operators $b_{2j}=c_{-1,j}$ and $b_{2j+1}=c_{+1,j}$, in which case $t,t^\prime$ are nearest and next-nearest neighbor hopping parameters, while $U,U_2$ are on-site and nearest neighbor interaction parameters, i.e. $H=t\sum_i(b^\dagger_{i+1}b_i+b^\dagger_ib_{i+1})-t^\prime\sum_i(b^\dagger_{i+2}b_i+b^\dagger_ib_{i+2})+\frac{U}{2}\sum_in_i(n_i-1)+U_2\sum_in_{i+1}n_i$.  Aside from the longer range hopping, Eq.~(\ref{eq:zigzag}) maps onto the model introduced by Dalla Torre {\it et al}. in considering polar molecules in optical lattices.  Unfortunately, based on their analysis, one expects that the Haldane phase is not stable when $U/U_2$ is large -- which is the physically relevant regime considered in \cite{Spielman}.  (Our numerics confirm this expectation.)  Here we argue that by using a Feshbach resonance \cite{feshbach}, one can reduce $U$, driving the system into the Haldane phase.  The lossy nature of bosonic Feshbach resonances aids us, as the quantum Zeno effect converts the resulting large 3-body recombination rate into a suppression of the probability of having more than two particles on any given site -- further stabilizing the Haldane phase.

\begin{figure}[t]
  \includegraphics[width=0.46\textwidth]{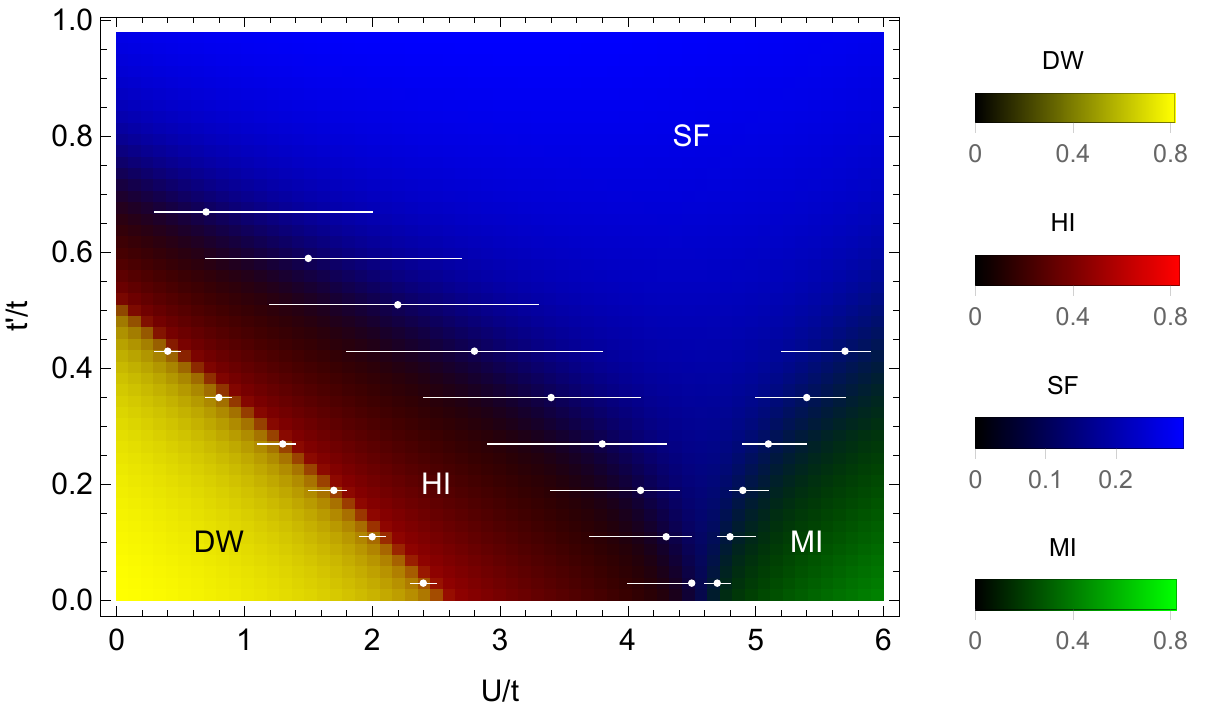}
  \caption{(Color Online) Representative slice of the phase diagram of the model in Eq.~(\ref{eq:zigzag}). Here $U_2/t=2.5$, and the maximum occupation of any site is 2. Dots show our best estimate of phase boundaries, as determined by bipartite number fluctuations, and lines represent error bars \cite{Sup}.  Yellow, Red, Blue, and Green show amplitude of the correlation functions in Eq.~(\ref{eq:cf}) at a separation of $s=80$ sites in a chain of length $256$.  At shorter lengthscales, correlations are of similar size, but the boundaries are less sharp.}
  \label{fig:fig2}
\end{figure}

More concretely, we consider the $F=1$, $m_F=1,0$ states of $^{87}$Rb. The coefficient $U$ is proportional to the scattering length associated with two atoms in the same magnetic sublevel.  As illustrated in Fig. \ref{fig:fig1}(b), this scattering length can be manipulated by applying a magnetic field. Near $B_1\sim 661.43$G there is a zero-crossing where the interactions between two $m_F=0$ atoms vanish, while near $B_2\sim 685.43$G there is a similar zero crossing for $m_F=1$ \cite{Marte}.  We envision rapidly switching the magnetic field between these two fields, as illustrated in Fig. \ref{fig:fig1}(c).  As long as the switching time is short compared to the other scales in the problem ($h/U,h/t\sim h/E_R\approx 0.27$ms for laser wavelength $\lambda=789$nm) the effective interaction in each spin channel will be given by time-averaging the instantaneous Hamiltonian $U_{\rm eff}=\int_0^t U(\tau)d\tau$ \cite{Blanes, Bukov, Eckardt}. Even though at any given time the interactions in the two channels will be different, this time averaged interaction is the same for each spin species, and $U$ will be the same on all sites.  This technique effectively halves the strength of the on-site interaction as in Fig. \ref{fig:fig1}(b).  The coefficient $U_2$ is largely unaffected. We find that one can achieve a ratio of on-site to nearest neighbor interaction of $U/U_2\approx 1.6$, for a lattice depth of $V_0=2E_R$ (for the effect of a higher band, see the Supplemental Material \cite{Sup}). By appropriately tuning the transverse confinement, one can take $U_2/t=2.5$, yielding $U/t=4$.  Figure~\ref{fig:fig2} shows the phase diagram for this model, revealing that these parameters place the system within the Haldane phase regime.

The zero crossings are very close to Feshbach resonances, and hence induce a large 3-body loss rate $K_3$.  In the present circumstance this is advantageous.  Following the logic in \cite{Zoller}, when $K_3$ is large, there is a strong suppression of the process in which a third particle hops onto a site containing two other particles.  This suppression can be modeled by a complex on-site three-body repulsion of strength $U_{3b}\sim -ihK_3 n^2/12$. We estimate that one can get $|U_{3b}|\sim 10E_R$ for a typical on-site particle density $n\sim 2.1\times 10^{15}$cm$^{-3}$ and typical on-resonance three-body rate $K_3\sim 10^{-25}$cm$^6$/s.  Since $|U_{3b}|$ is larger than the other scales in the problem, it can be replaced by a constraint that no more than two particles can occupy any site.

With this constraint we use the DMRG to calculate the properties of the model in Eq.~(\ref{eq:zigzag}).  We start with a infinite DMRG algorithm to grow the system to desired size, and then do finite DMRG sweeps until we reach convergent. This technique is typically understood as systematically optimizing a variational wavefunction in the form of a matrix product state \cite{Schollwock}.  The degree of approximation is controlled by the bond dimension $d$.  We have considered systems as large as $L=512$ sites, and bond dimensions as large as $d=500$.  The algorithm is more efficient if we alter the boundary conditions to break the potential four-fold degeneracy of the Haldane-phase groundstate, and the potential two-fold degeneracy of the density wave groundstate.  In particular we used boundary conditions which pin a vacancy at the left-most site, and doublon at the right-most site.  We analyze convergence with bond dimension and system size in the supplementary information.  From these studies we expect that experiments on systems of size $L\sim 60$ will see significant finite-size effects near the phase boundaries, but the bulk physics is unchanged, and such experiments will be able to unambiguously observe all of the relevant physical phenomena.

The order in the Haldane, Mott insulator, density wave phases are encoded in string (str), parity (MI) and density wave (DW) correlation functions \cite{Nijs, Altman2}
\begin{align}\label{eq:cf}
C^{\rm str}_{ij}&=\left<\delta n_ie^{i\pi\sum_{i<k<j}\delta n_k}\delta n_j\right>,\nonumber\\
C^{\rm MI}_{ij}&=\left<e^{i\pi\sum_{i\le k\le j}\delta n_k}\right>,\nonumber\\
C^{\rm DW}_{ij}&=(-1)^{j-i}\left<\delta n_i\delta n_j\right>,
\end{align}
where $\delta n_k=n_k-1$.  The phase factor $e^{i\pi\sum_{i<k<j}\delta n_k}=\pm 1$ depends on if the number of quasiparticles between sites $i$ and $j$ is even or odd. In the superfluid phase all of the three correlation functions fall to zero as $i$ and $j$ are separated.  In the Haldane/Mott insulator phase, only $C^{\rm str}/C^{\rm MI}$ has long-range order, while in the density wave phase all the three correlation functions are nonzero.

\begin{figure}[t]
  \includegraphics[width=0.45\textwidth]{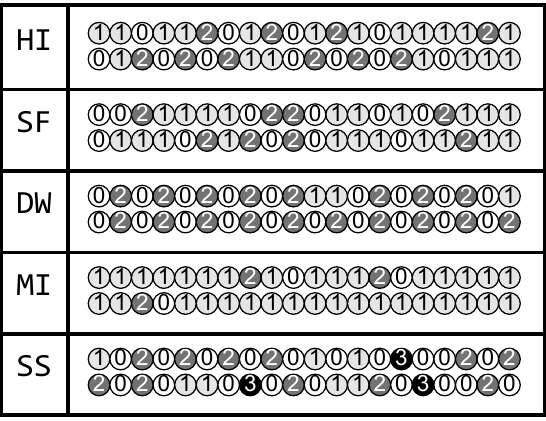}
  \caption{Typical configurations of occupation numbers extracted from the central
   20 sites of our DMRG wavefunctions, modeling single-shot quantum gas microscope images.  Configurations correspond to parameters in which different forms of order can be observed: HI($U/t=4$, $U_2/t=2.5$ for $t^\prime/t=0$), SF($U/t=4$, $U_2/t=2.5$ for $t^\prime/t=1$), DW($U/t=1$, $U_2/t=2.5$ for $t^\prime/t=0$), MI($U/t=10$, $U_2/t=2.5$ for $t^\prime/t=0$).  The last figure has short-range supersolid correlations: SS($U/t=0$, $U_2/t=2.5$ for $t^\prime/t=0.6$ for maximum occupation number 3). Each circle resembles a single site, and the number in the circle tells how many atoms are on this site. We show two independent realizations for each phase.}
  \label{fig:fig3}
\end{figure}

We additionally study the single particle density matrix $C^{\rm SF}_{ij}=\langle b_i^\dagger b_j\rangle$ and the bipartite number fluctuations, $D_j =\langle N_{i<j}^2\rangle-\langle N_{i<j}\rangle^2$, where $N_{i<j}=\sum_{i=1}^{j-1} n_i$ is the number of particles to the left of site $j$.  The superfluid phase is characterized by power-law behavior of the density matrix, and enhance bipartite number fluctuations when compared to the incompressible insulating phases.  We found that these number fluctuations were the most reliable way to extract the phase boundaries between the superfluid and insulating phases \cite{Sup}.  In particular, due to the different scaling with system size, the number fluctuations in half the chain, $D_{L/2}$, form plateaus in each of the phases, and the phase boundaries correspond to peaks in the slope $d D_{L/2}/dt^\prime$.  We use the full width half max of these peaks as an estimate of the accuracy of these boundaries.  This approach is adapted from \cite{fluckt}, and is similar to finding phase boundaries from peaks in specific heat.

Additionally, the DW to HI transition can be accurately determined from the properties of $C^{\rm DW}$.  Applying finite size scaling  \cite{fsc} to the  asymptotic behavior of this correlation function yields a DW-HI boundary which agrees with our calculation using the number fluctuations.  The various superfluid-insulator transitions are not amenable to this standard finite size scaling analysis:  they have behavior related to Kosterlitz-Thouless transitions, and are harder to determine.  In addition to our technique of looking at number fluctuations, these transitions can be identified by looking at the excitations spectrum \cite{Rossini} or superfluid stiffness \cite{Stiff}, by taking moments of $C^{\rm SF}$ \cite{pai,roomany}, or by comparing the power law decay of $C^{\rm SF}$ to a Luttinger liquid model \cite{pow,minguzzi}.  More discussion of the critical behavior can be found in \cite{kurdestany}.  Due to the significant finite size effects, it is unlikely that an experiment would be able to accurately determine these phase boundaries.

The correlation functions in Eq.~(\ref{eq:cf}) are directly measurable via a quantum gas microscope \cite{Ott, Bakr, Bakr2, Sherson, Endres}.  One projects the quantum state into one in which there is a definite number of particles on each site -- giving a single realization of $\{n_i\}$.  Repeating the measurement many times allows one to extract the expectation values in Eq.~(\ref{eq:cf}).  This technique has already been used to measure the parity order \cite{Endres}.  

In addition to showing the phase boundaries, Fig.~\ref{fig:fig2} shows the size of correlations on a length-scale of 80 sites.  In a significant part of the phase diagram, the string correlations are large but all other correlation functions vanish.  This corresponds to the desired Haldane phase.  The $HI$, $DW$, and $MI$ correlations at shorter lengths scales are of similar strength, but display less sharp boundaries.  As would be expected, the $SF$ correlations are strongly length-dependent. 
They are also extremely hard to measure in an experiment.
 The simplest experimental knob for moving through this phase diagram is the strength of the Raman beams, which changes $t$ while leaving all other scales unchanged.  Additionally, both $t$ and $t^\prime$ are exponentially sensitive to the lattice depth, while the ratio between $U$ and $U_2$ is controlled by modifying the time dependence of the magnetic field.

In addition to exploring the expectation values of the various correlation functions, we use a novel Monte-Carlo sampling algorithm to stochastically generate `typical' cold-gas microscope images \cite{Sup}.  Given the DMRG wavefunction $|\psi\rangle$, we first calculate the probability that site-1 had $0,1$ or $2$ particles on it.  We use these probabilities to choose one of these sectors, and project the wavefunction into that sector.  This calculation is then repeated on site-2, using the new wavefunction...  Figure~\ref{fig:fig3} shows  configurations generated by this algorithm, which should be representative of what is seen in an experiment.  We emphasize that these are not cartoons, but rather are unbiased samples.  As expected, in the HI phase the doublons and holons alternate, with a variable number of singly-occupied sites between them.  This can be contrasted with the SF phase, where there is no ordering of the doublons and holons.  In the DW phase, doublons and holons alternate.  In these images one sees a small number of defects in the order -- as should be expected.  In the MI phase the images show very few holons and doublons -- and those which exist are tightly bound together.  In this figure we also show images with supersolid (SS) correlations that can appear when we relax the constraint forbidding double occupancy.  The physics of this regime will be discussed below.

\begin{figure}[t]
  \includegraphics[width=0.46\textwidth]{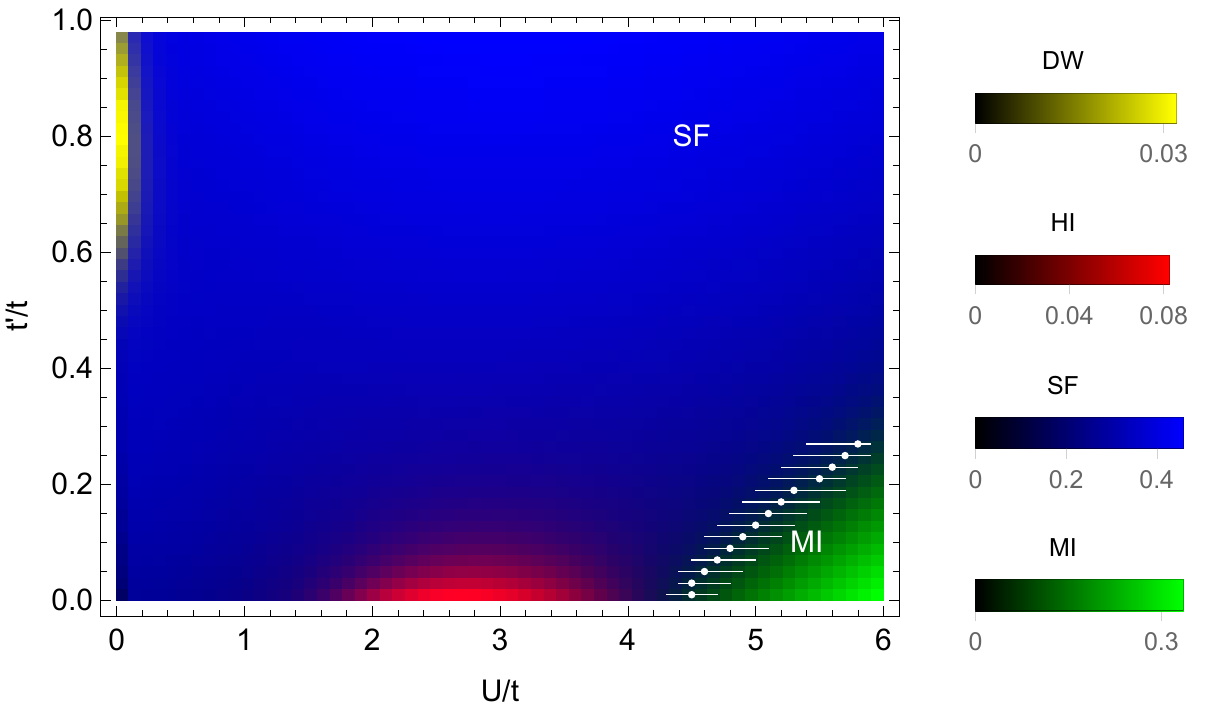}
  \caption{(Color Online) Representative slice of phase diagram when maximum occupation of any site is 3.  All parameters and symbols same as Fig.~\ref{fig:fig2}.  Note change of scale on color bars.  For these parameters there are regions with short or intermediate-range density wave (yellow) or Haldane order (red), but  no long-range order. }
  \label{fig:fig5}
\end{figure}

To illustrate the role of the three-body constraint, we repeated our calculations, allowing the on-site particle number to be as large as 3.  Figure \ref{fig:fig5} shows the analog of Fig. \ref{fig:fig2}.  All correlations, except those corresponding to SF order, are much weaker.  Short and medium range HI, MI, and DW correlations are detectable, but our scaling analysis suggest that for these parameters there is no long-range DW or HI order.

 For small on-site interaction $U$ and finite next nearest hopping $t'$, we find a superfluid region with short-range density wave order,  which is suggestive of proximity to a supersolid phase (SS). Such a phase would be more familiar in the language of the ``zig-zag" ladder picture:  The atoms form a superfluid which preferentially sits on one leg of the ladder.  An alternative cartoon can be constructed from the DW state ``2020202020."  Because of the next-nearest neighbor hopping, one can produce a triplon-singlon pair ``2020103020," and these defects may be mobile.  Forbidding triple occupancy eliminates these excitations, and prevents the occurrence of this phase in the constrained model.  In addition to such triplons and singlons, the configurations in Fig.~\ref{fig:fig3} display defects where atoms have hopped from even to odd sublattices.  These defects are responsible for the short-range nature of the correlations.

To summarize, we have proposed a way to realize the Haldane phase in a gas of $^{87}$Rb atoms trapped in a zigzag optical lattice, where a different atomic spin state is trapped on each leg of the ladder.  One reduces the on-site interactions (relative to the nearest neighbor interactions) by rapidly sweeping the magnetic field between two zero-crossings associated with Feshbach resonances in each of the spin states.  The proximity to the Feshbach resonances introduces large three-body loss, which via the quantum Zeno effect prevents triple-occupation.  We calculate the phase diagram of this model, and find that the Haldane phase is experimentally realizable. We modeled a quantum-gas microscope experiment, and found that one can readily identify the string order of the Haldane phase in individual images.  More quantitative tests require averaging over several images.  Such averaging has been used to identify other non-local order parameters \cite{Endres}.  We further show that without the constraint on particle number, this model shows hints of a supersolid phase (cf. \cite{Esslinger, Ketterle}).  One would need to use other techniques, however, to experimentally reach this supersolid regime. Seeing the string order in the Haldane phase would be a remarkable triumph in engineering quantum matter.

J.X. acknowledges helps from Matthew Reichl on the numerical calculation. In addition to our custom-built DMRG software, we performed many of the calculation using the ITensor library (http://itensor.org). E.J.M. acknowledges help from Joshua Squires on aspects related to the phase boundaries. This research is supported by the NBRPC (No. 2013CB922002), NNSFC (No. 11504021, 11574028), NSF (No. PHY-1508300), and FRFCU (No. FRF-TP-17-023A2).

\clearpage
\onecolumngrid

\begin{center}
\textbf{\large Supplementary Material: Realizing the Haldane Phase with Bosons in Optical Lattices}
\end{center}

\setcounter{equation}{0}
\setcounter{figure}{0}
\setcounter{table}{0}
\setcounter{page}{1}
\renewcommand{\thepage}{S\arabic{page}}  
\renewcommand{\thesection}{S\arabic{section}}   
\renewcommand{\thetable}{S\arabic{table}}   
\renewcommand{\thefigure}{S\arabic{figure}}
\renewcommand{\theequation}{S\arabic{equation}}
\renewcommand{\bibnumfmt}[1]{[S#1]}
\renewcommand{\citenumfont}[1]{S#1}

\section{DMRG Algorithm}\label{rev}
To study our effectively one-dimensional system we use the Density Matrix Renormalization Group (DMRG) method.  There are a number of excellent reviews of the technique \cite{Schollwock} and pedagogical sample code which makes it easier to understand the mechanics \cite{codeexamples}.  Here we give a quick introduction, and elaborate on our explanation of how we extract cold-gas microscope configurations, as in Fig. 3 of the main text.

\subsection{Configurations and blocks}
The DMRG algorithm involves sequentially repartitioning the system into configurations of different shape, refining a variational wavefunction as one sweeps through the configurations. For a chain of length $L$, we consider $L-3$ different configurations.  In the $j$'th configuration, the left block contains $j$ sites, the right block contains $L-j-2$ sites, and two extra sites -- labeled $a$ and $b$ -- lie between them.  The basis for the left block is $|1\rangle_L^j,|2\rangle_L^j\cdots |d_L\rangle_L^j$, for the right block is $|1\rangle_R^j,|2\rangle_R^j\cdots |d_R\rangle_R^j$, and the central two sites are $|1\rangle_a^j\cdots |m\rangle_a^j$ and 
$|1\rangle_b^j\cdots |m\rangle_b^j$. Here $m$ encodes the size of the Hilbert space for a single site: $m=3$ or 4 when we truncate to 2 or 3 particles per site.  In the absence of any approximations $d_L= m^j$ and $d_R=m^{L-j-2}$. 

The bases for the various partitions are related by tensors $\Gamma^L$ and $\Gamma^R$,
\begin{equation}
|i\rangle_L^j = \sum_{k\alpha} \Gamma^{Lj}_{ik\alpha} |k\rangle_L^{j-1}|\alpha\rangle_a^{j-1},\quad
|i\rangle_R^j = \sum_{k\beta} \Gamma^{Rj}_{ik\beta} |k\rangle_R^{j+1}|\beta\rangle_b^{j+1}.
\end{equation}
Here $i$ indexes the basis states of the Left/Right block in partition $j$, while $k$ runs over the 
basis in partition $j-1$ or $j+1$, while $\alpha$ and $\beta$ index the basis states for the $a$ and $b$ site in partition $j$.
In the absence of approximations, the tensors are unitary,
\begin{equation}
\sum_{k\alpha}  \Gamma^{Lj}_{ik\alpha}  (\Gamma^{Lj}_{i^\prime k\alpha} )^* =\delta_{i i^\prime},\quad
\sum_{i}  \Gamma^{Lj}_{ik\beta}  (\Gamma^{Lj}_{i k^\prime\beta^\prime} )^* =\delta_{kk^\prime} \delta_{\beta\beta^\prime},
\end{equation}
and similar with $L\to R$.

A wavefunction in the $j$'th partition, can be written as a rank 4 tensor of size $d_L\times d_R\times m\times m$,
\begin{equation}
|\psi\rangle = \sum_{\ell r\alpha\beta}\psi^{(j)}_{\ell r\alpha\beta} |\ell\rangle_L^j |r\rangle_R^j |\alpha\rangle_a^j |\beta\rangle_b^j.
\end{equation}
Using the relationship between the basis functions, one has 
\begin{equation}
\psi^{(j+1)}_{\mu\nu\beta t} = \sum_{ \ell r \alpha} \psi^{(j)}_{\ell r \alpha \beta} \Gamma^{R j}_{r\nu t} \left( \Gamma^{L (j+1)}_{\mu\ell\alpha}\right)^*.
\end{equation}


\subsection{Approximation}
In the DMRG one truncates the basis used for each block, keeping only the $d$ ``most important" basis states, where $d$ is referred to as the bond dimension.  The ``important" states correspond to the $d$ largest eigenvalues of the reduced density matrix for that block.  For example, given a wavefunction
$\psi^{(j)}_{\ell r\alpha\beta}$, one can construct a reduced density matrix for the left block by tracing out the other degrees of freedom,
\begin{equation}
\rho^{Lj}_{\ell \ell^\prime}= \sum_{r\alpha\beta} \psi^{(j)}_{\ell r\alpha\beta}\left(\psi^{(j)}_{\ell^\prime r\alpha\beta}\right)^*.
\end{equation}
The eigenvalues of this matrix are denoted $\lambda_i$, and $\sum_i \lambda_i=1$. 

Rather than storing the basis vectors, in the DMRG one stores matrix elements of the various operators which are needed to construct the Hamiltonian.  One only stores operators which act on a single block -- which in the truncated basis requires at most $d^2$ numbers.  

The DMRG is a systematic way of generating a sequence of truncations.  In the next few paragraphs we explain how one generates a new, improved truncation from a prior non-optimal truncation.

Consider the $j$'th configuration of the non-optimal truncation.  One uses the stored information to generate the Hamiltonian as a $(d^2m^2)\times(d^2 m^2)$ matrix.  One finds the eigenvector of this matrix corresponding to the smallest eigenvalue, which yields an approximate $\psi^{(j)}_{\ell r\alpha\beta}$.  One then finds the reduced density matrix associated with the leftmost $j+1$ sites,
\begin{equation}
\rho^{L(j+1)}_{\ell \alpha ,\ell^\prime\alpha^\prime}= \sum_{r\beta} \psi^{(j)}_{\ell r\alpha\beta}\left(\psi^{(j)}_{\ell^\prime r\alpha^\prime\beta}\right)^*,
\end{equation}
where the $m\times d$ pairs $(\ell,\alpha)$ provide indices for the basis state of these $j+1$ sites.  One then finds the $d$ largest eigenvalues, and the corresponding wavefuctions,
$\phi^s_{\ell\alpha}$, with $s=1,2,\cdots d$.  These become the basis states
 for left block of the $j+1$'st configuration.  These states also act as the transformation matrix 
 \begin{equation}
 \Gamma^{L(j+1)}_{s\ell\alpha}=\phi^s_{\ell\alpha},
 \end{equation}
 and are used to transform the matrix elements of the operators which either act on the left block, or the $a$ site.  These new matrix elements are stored.  Note, the constructed $\Gamma$ tensors are no longer unitary, but rather act as projection operators.  This new truncation is better than the old.
 
 The process can then be repeated, using the matrices from the $j+1$'st configuration to update the left-block matrices for the $j+2$'nd configuration.  One ``sweeps" from left to right until all configurations are updated.  Sweeping right to left then updates the right-block matrices.  These sweeps are repeated until a fixed point is reached.
 
 There are several tricks to speed up the calculation.  First, symmetries and conservation laws (such as number conservation in our problem) imply that the $(d^2m^2)\times(d^2 m^2)$ Hamiltonian matrix is block diagonal.  One need only diagonalize the block corresponding to the physical quantum numbers.  Second, one can use iterative algorithms for the diagonalization.  The wavefunctions from prior steps in the sweep can be used to find good starting points for the iterations.  As emphasized in the main text, appropriate boundary conditions can also eliminate degeneracies or near degeneracies, and accelerate convergence.
 
 The approximate wavefunction generated by this technique is a matrix-product-state, and the DMRG algorithm can be interpreted as a means of finding the optimal variational matrix-product-state wavefunction \cite{mps}.
 
 \subsection{Cold gas microscopy}
In cold atoms one has a unique probe where one counts how many particles are on each lattice site \cite{Ott, Bakr, Bakr2, Sherson, Endres}.  This is a projective measurement which not only tells one about the average occupation of sites, but also the fluctuations and their correlations.  

All measurement processes are stochastic, and in each run of the experiment one will see a different image in the microscope.
We can use the information stored in our DMRG calculation to stochastically generate configurations which are drawn from the same distribution as the experimental images.

In particular, we need the following:  The set of transformation matrices,  $\Gamma^{Lj}_{\ell k\alpha}$, $\Gamma^{Rj}_{r k\alpha}$,  the wavefunction in the first configuration $\psi^{(1)}_{\ell r \alpha \beta}$, and matrix elements for each of the operators which project into the space where a given number of particles sit on a given site.  We let $\hat P^k_n$ be the operator that projects into the space where the $k$'th site has exactly $n$ particles.  
If $2\leq k\leq L-2$, then this operator acts on the $a$ site in the $j=k-1$'th configuration.  If $k=1,L-1$, or $L$, it respectively acts on the left block of the $j=1$'th configuration, the $b$ site of the $j=L-3$ configuration, and the right block of the $j=L-3$ configuration.   
 
We first calculate the probabilities that the first site has $n$ particles on it -- for $n=0,1,...n_{\rm max}$, with $n_{\rm max}=2$ or $3$, depending on the calculation.  These probabilities are extracted from the wavefunction via
 \begin{equation}
P_{n_1=n}= \sum_{r\alpha\beta} \sum_{\ell \ell^\prime} (\psi^{(1)}_{\ell r \alpha \beta})^* \psi^{(1)}_{\ell^\prime r \alpha \beta} (P^1_n)_{\ell \ell^\prime}.
 \end{equation}
 We then use a random number generator to choose one of these options with the appropriate probability.  Denoting our choice $n_1$, we project our wavefunction into that space via
 \begin{equation}
 \psi^{(1)}_{\ell r \alpha \beta} \to \sum_{\ell^\prime}  (P^1_{n_1})_{\ell \ell^\prime}  \psi^{(1)}_{\ell^\prime r \alpha \beta}.
 \end{equation}
 We then use this new wavefunction to calculate the probability for the various values of $n_2$, conditioned on the measured $n_1$.  For notational simplicity, we suppress reference to the prior measurements, writing,
  \begin{equation}
P_{n_2=n}= \sum_{r\ell\beta} \sum_{\alpha \alpha^\prime} (\psi^{(1)}_{\ell r \alpha \beta})^* \psi^{(1)}_{\ell r \alpha^\prime \beta} (P^2_n)_{\alpha \alpha^\prime}.
 \end{equation}
 Again we use a random number generator to choose one of these options with the appropriate probability, and project via
  \begin{equation}
 \psi^{(1)}_{\ell r \alpha \beta} \to \sum_{\alpha^\prime}  (P^2_{n_2})_{\alpha \alpha^\prime}  \psi^{(1)}_{\ell r \alpha^\prime \beta}.
 \end{equation}
 Next we use the transformation matrices to convert this wavefunction to the next sector,
   \begin{equation}\label{trans}
 \psi^{(2)}_{\ell r \alpha \beta}=
 \sum_{ \ell^\prime r^\prime \alpha^\prime} \psi^{(1)}_{\ell^\prime r^\prime \alpha^\prime \alpha} \Gamma^{R 1}_{r^\prime r \beta} \left( \Gamma^{L 2}_{\ell\ell^\prime \alpha^\prime}\right)^*.
 \end{equation}
 We then use this wavefunction to calculate the probabilities for various values of $n_3$, conditioned on the prior measurements.  One continues in this manner until all occupation numbers are determined.
 
The reason that we shift between the configurations is so that our projectors always act on blocks which are not truncated, and therefore introduce no approximations.
 One may be concerned that with our truncated basis, transformations such as Eq.~(\ref{trans}) are not unitary.  We verify that the norm of our wavefunction does not decrease as long as we have a sufficiently large bond dimension.   Physically, the ultimate state that we project into is ``close enough" to the grounds state that it can be well-approximated using the truncated basis.

\section{Higher Band Effects}
In our approach, the Haldane phase appears at relatively low lattice depths, $V_0\sim 2E_R$.  
Here we model higher band effects, and show that they are negligible.

\begin{figure}[h]
  \includegraphics[width=0.7\textwidth]{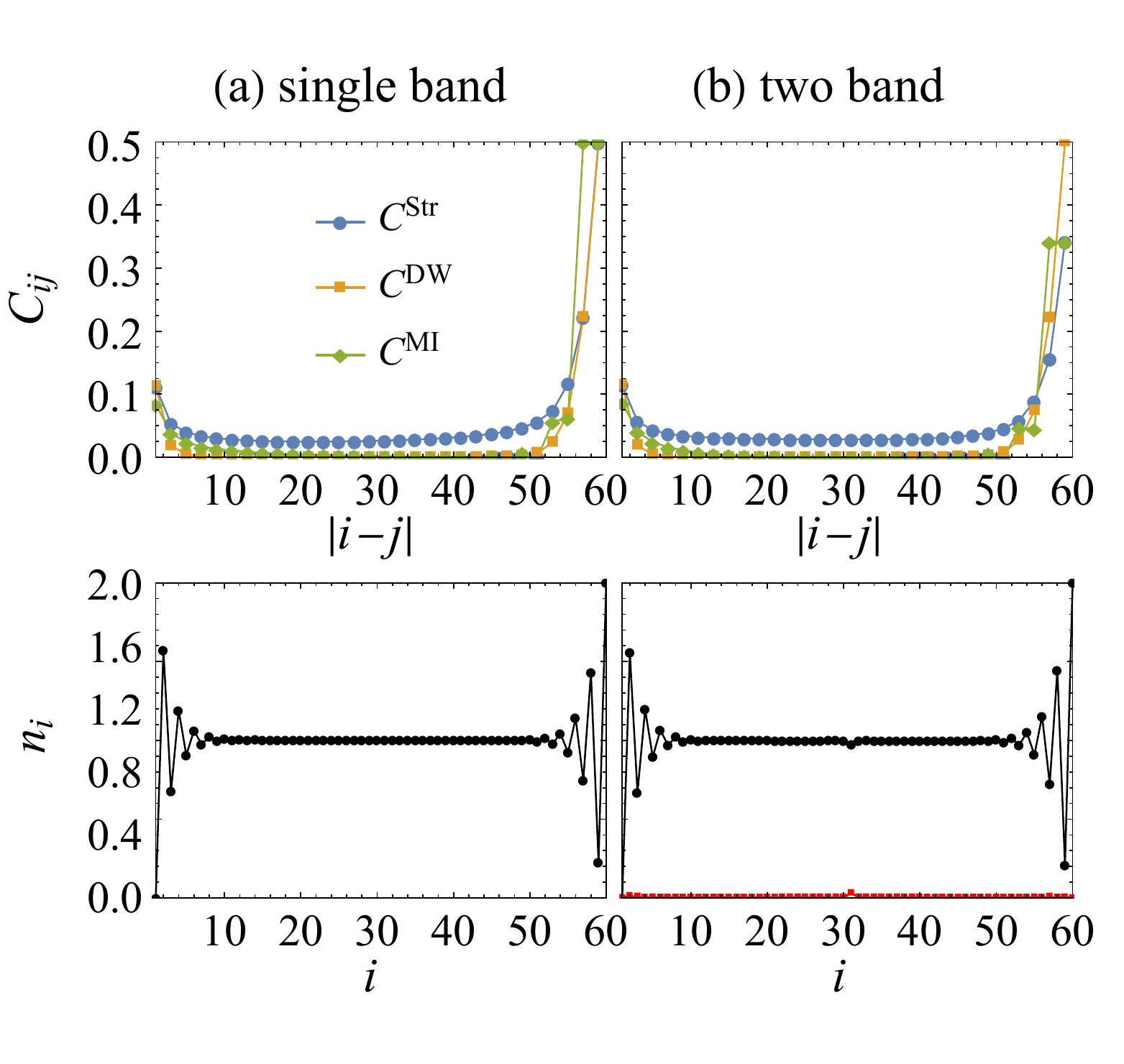}
  \caption{Upper panel: correlation functions and lower panel: particle densities for (a) single band and (b) two band Bose-Hubbard model. The parameters are chosen by taking 
  $t'/t=0.3$, 
 $U_2/t=2.5$, 
 and $V_0=2 E_R$, which then specifies all other quantities (see the text). 
 The occupancy of the second band (red line) is extremely small.}
  \label{fig:figs1}
\end{figure}

Using similar notation as in the main text, we consider a two-band tight binding model,
\begin{align}\label{eq:twoband}
H&=t\sum_i(b^\dagger_{i+1}b_i+b^\dagger_ib_{i+1})-t'\sum_i(b^\dagger_{i+2}b_i+b^\dagger_ib_{i+2})+\frac{U}{2}\sum_in_i(n_i-1)+U_2\sum_in_{i+1}n_i\nonumber\\
&+\bar{t}\sum_i(\bar{b}^\dagger_{i+1}\bar{b}_i+\bar{b}^\dagger_i\bar{b}_{i+1})-\bar{t}'\sum_i(\bar{b}^\dagger_{i+2}\bar{b}_i+\bar{b}^\dagger_i\bar{b}_{i+2})+\frac{\bar{U}}{2}\sum_i\bar{n}_i(\bar{n}_i-1)+\bar{U}_2\sum_i\bar{n}_{i+1}\bar{n}_i\nonumber\\
&+2\tilde{U}\sum_in_i\bar{n}_i+\tilde{U}_2\sum_i(n_{i}\bar{n}_{i+1}+\bar{n}_in_{i+1})+\frac{\tilde{U}}{2}\sum_i(\bar{b}^\dagger_i\bar{b}^\dagger_ib_ib_i+b^\dagger_ib^\dagger_i\bar{b}_i\bar{b}_i)\nonumber\\
&+\lambda\sum_i(b^\dagger_{i+1}\bar{b}_i+\bar{b}^\dagger_ib_{i+1}-\bar{b}^\dagger_{i+1}b_i-b^\dagger_i\bar{b}_{i+1})+\delta\sum_i\bar{n}_i,
\end{align}
where the bars label the second band, and the tilde labels the interaction between these two bands. These coefficients are connected with the Wannier functions for the two bands $w_1(x)$ and $w_2(x)$ via \cite{Spielman}
\begin{gather}
t=\Omega\int w_1^*(x-a/4)w_1(x+a/4)dx,\quad \bar{t}=\Omega\int w_2^*(x-a/4)w_2(x+a/4)dx,\\
t'=-\int w_1^*(x)H_0w_1(x+a)dx,\quad \bar{t}'=-\int w_2^*(x)H_0w_2(x+a)dx,\\
U=U_0\int|w_1(x)|^4dx,\quad \bar{U}=U_0\int|w_2(x)|^4dx,\\
U_2=U_0\int|w_1(x)|^2|w_1(x+a/2)|^2dx,\quad \bar{U}_2=U_0\int|w_2(x)|^2|w_2(x+a/2)|^2dx,\\
\tilde{U}=U_0\int|w_1(x)|^2|w_2(x)|^2dx,\quad \tilde{U}_2=U_0\int|w_1(x)|^2|w_2(x+a/2)|^2dx,\\
\lambda=\Omega\int w_1^*(x-a/4)w_2(x+a/4)dx,\\
\delta=\int w_2^*(x)H_0w_2(x)dx-\int w_1^*(x)H_0w_1(x)dx,
\end{gather}
where $H_0=p^2/(2m)+V_0\cos(2k_Lx)/2$ is the Hamiltonian of one spin component in the absence of the Raman laser with Rabi frequency $\Omega$.  The laser recoil energy is $E_R=\hbar^2k_L^2/(2m)$,  and $U_0$ denotes the reduced interaction energy, related to the scattering length and the transverse confinement. We write the Wannier functions in terms of Mathieu functions, and numerically perform the integrals. 

We fix $t'/t=0.3$, $U_2/t=2.5$, and $V_0=2E_R$.  We then obtain $\bar{t}/t=-1.00$, $\bar{t}'/t=1.07$, $U/t=3.95$, $\bar{U}/t=2.53$, $\tilde{U}/t=1.91$, $\bar{U}_2/t=2.48$, $\tilde{U}_2/t=3.45$, $\lambda/t=0.70$, $\delta/t=4.84$, $\Omega/t=4.23$, $U_0/t=10.00$.   

Of particular note is the fact that for these parameters, the interaction between neighboring atoms in the same band ($U_2$ and $\bar{U}_2$) are smaller than the interactions between atoms in different bands,  $\tilde{U}_2$.  This feature has to do with the shape of the Wannier states for each bands, and will tend to discourage states where there is occupation of both bands.  

We carry out DMRG simulations of this two-band model, using the same boundary conditions as in our single band calculations.  We find results which are nearly indistinguishable from our one-band model.  To further illustrate this equivalence, we extract order parameters
\begin{align}\label{eq:cf}
C^{\rm str}_{ij}&=\left<\delta \tilde{n}_ie^{i\pi\sum_{i<k<j}\delta \tilde{n}_k}\delta \tilde{n}_j\right>,\nonumber\\
C^{\rm MI}_{ij}&=\left<e^{i\pi\sum_{i\le k\le j}\delta \tilde{n}_k}\right>,\nonumber\\
C^{\rm DW}_{ij}&=(-1)^{j-i}\left<\delta \tilde{n}_i\delta \tilde{n}_j\right>.
\end{align}
Here the particle number is defined as the total number $\tilde{n}_i=n_i+\bar{n}_i$.
Our two-band results are shown on the right in Fig. \ref{fig:figs1} with the single band results on the left. Despite the relatively shallow potential, these results are nearly identical.  

Previous studies of higher band effects have seen similar results.  For example, Xu {\it et al.} used a Quantum Monte-Carlo approach to solve the problem of one-dimensional (1D) interacting Bosons in a 1D optical lattice \cite{xu}.  Although they found that the relationship between the scattering length and the Hubbard $U$ was renormalized at small lattice depths, the single band Bose-Hubbard model remained predictive in this regime.  A recent review of extended Hubbard models contains some discussion of multi-band effects \cite{dutta}.

\begin{figure}[t]
  \includegraphics[width=1\textwidth]{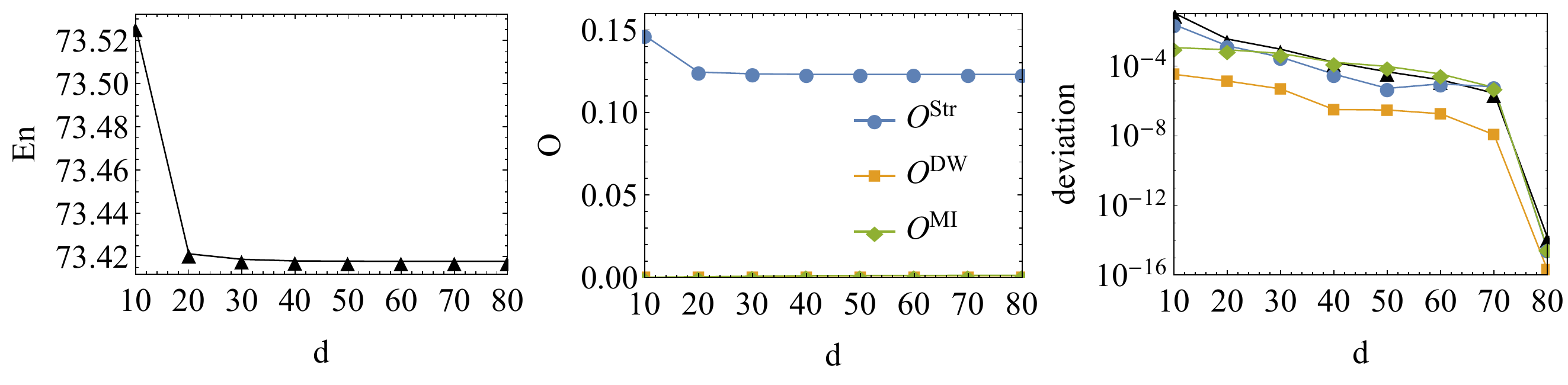}
  \caption{Convergence with bond dimension $d$ for the Haldane phase with maximum on-site occupation number 2. We vary $d$ from 10 to 500 but only show the results of $d<80$ here. We measure the deviation  of each quantity  relative to their value at d=500. The black triangles show the energy, while the blue circles/green diamonds/orange squares are the string/parity/density correlation functions, measured over the middle 30 sites on a chain with $L=60$:  $O= C_{i=L/4,j=3L/4}$. We use parameters $t'/t=0.1$, $U/t=4.0$, $U_2/t=2.5$.}
  \label{fig:figs2a}
\end{figure}

\begin{figure}[t]
  \includegraphics[width=1\textwidth]{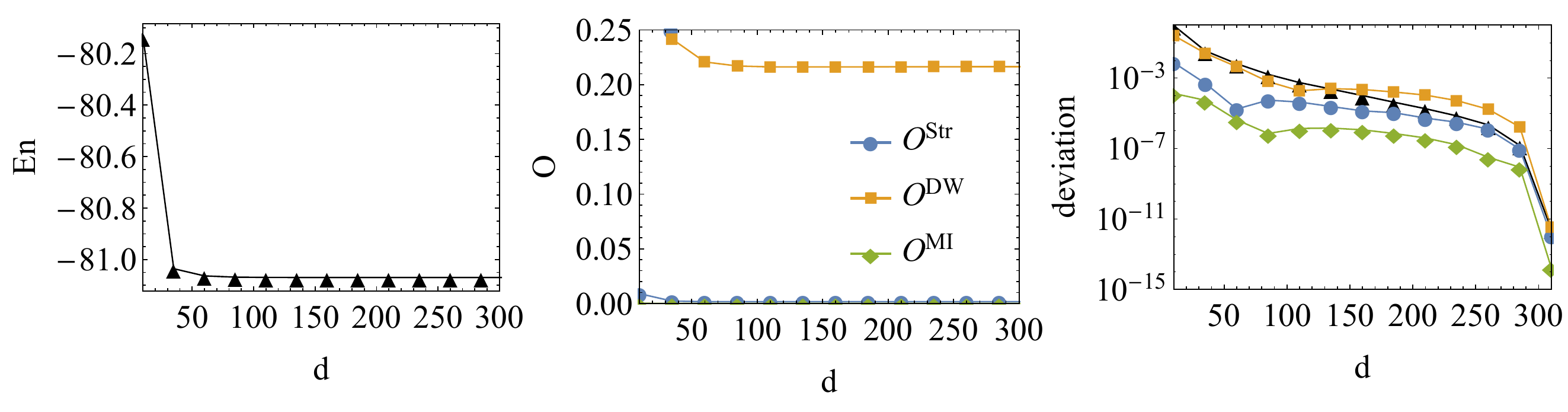}
  \caption{Convergence with bond dimension $d$ for the superfluid phase, in a regime with medium range density wave correlations.   We vary bond dimension $d$ from 10 to 500 and only show the results of $d<300$ here. Deviations are shown relative to the value of each quantity at d=500. We use parameters $t^\prime/t=0.6$, $U/t=0$, $U_2/t=2.5$,  and restrict the maximum on-site occupation number to be 3. The colors are the same as Fig. \ref{fig:figs2a}.  As expected for a gapless phase, the convergence rate depends on the system size.}
  \label{fig:figs2b}
\end{figure}

\section{Convergence Study with bond dimension}
As discussed in section~\ref{rev}, the DMRG can be interpreted as a variational technique, which rapidly optimizes the parameters in a matrix-product-state wavefunction.  The size of the matrices used  is called the bond-dimension $d$.  Physically $\log(d)$ is the maximum amount of entanglement entropy which is captured by this ansatz.  This technique is exact in the limit $d\to\infty$.   We take chains of length $L=60$, and vary the bond dimension from $d=10$ to $500$.   Correlation functions are measured over the central 30 sites.  For the calculations with large bond dimensions we use the ITensor package, as it is more efficient than the custom-built code that we used for many of our studies.

Figure~\ref{fig:figs2a} uses parameters corresponding to the Haldane Insulator phase.  This is a gapped phase, where the entanglement entropy is expected to be modest, and independent of system size.   Consequently, convergence is extremely rapid.  Figure~\ref{fig:figs2b} uses parameters corresponding to a region of the superfluid phase which displays strong medium-range density wave correlations.   This phase is gapless, and one expects more modest convergence with bond dimension.  Moreover the entanglement entropy should grow with system size,   Nonetheless the results appear to be quite accurate even for relatively small $d$.

%
%
%
%
%
%
%

\section{Convergence studies with system size, and extracting phase boundaries}
\begin{figure}[t]
  \includegraphics[width=0.6\textwidth]{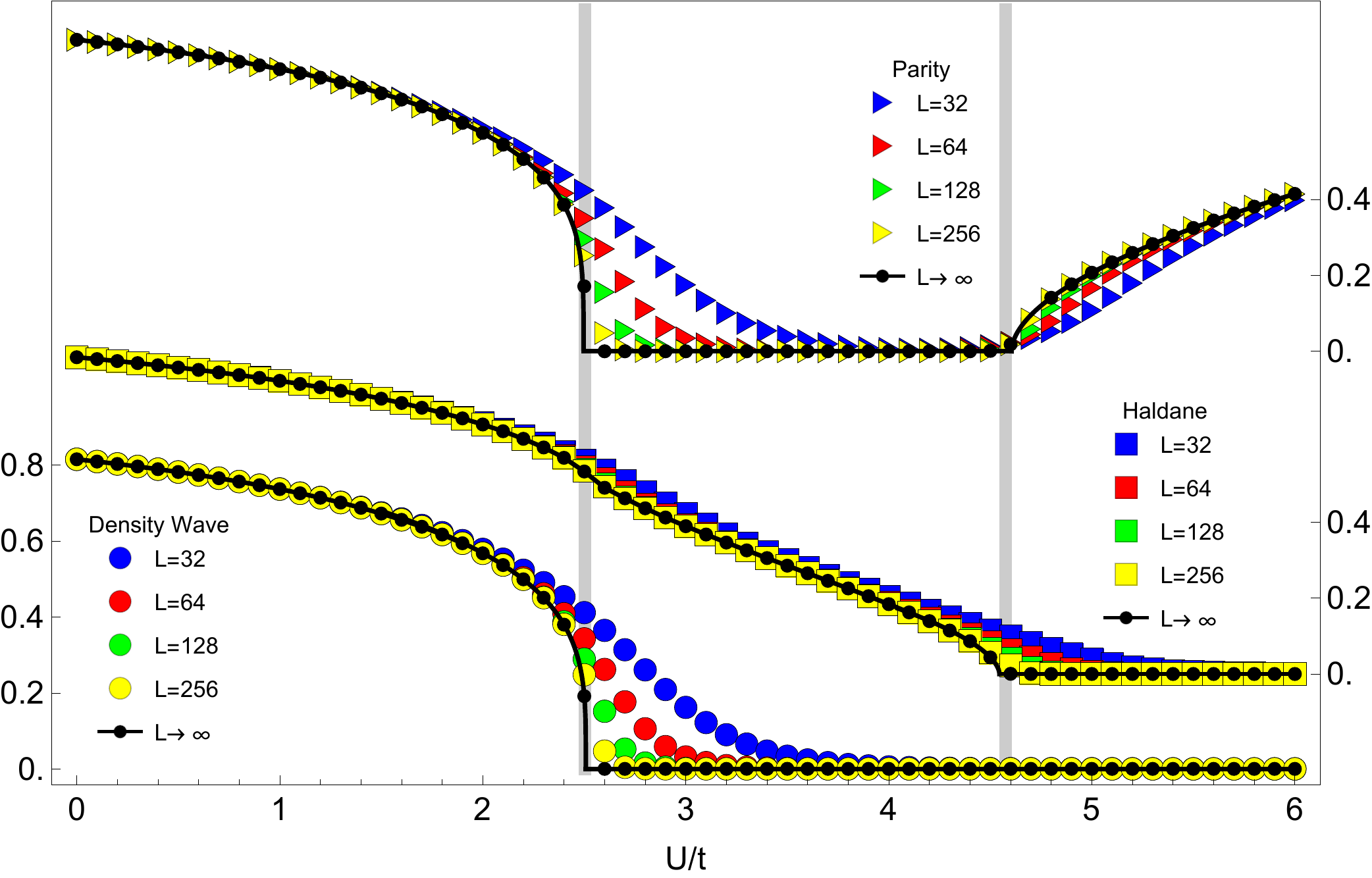}
  \caption{Correlation functions for $t^\prime=0$ and $U_2/t=2.5$.  Each plot shows the correlation function $O=C_{i,j}$ between sites $i=L/4$ and $j=3 L/4$ for $L=32,64,128,256$, and our extrapolation to $L=\infty$.  Vertical lines show the phase boundaries between the density wave, Haldane insulator, and Mott insulator (left to right).  As is clear from the scales, the vertical axes for the different correlation functions are offset from one-another. }
  \label{tp0scale}
\end{figure}

To understand finite size effects in our simulations, we repeat our calculations for different length chains $L$.  We calculate the correlation functions $O_L=C_{i=L/4,j=3L/4}$.  When $t^\prime=0$ and $U_2/t=2.5$ we find that we can readily extrapolate to the thermodynamic limit, modeling $O_L= O_\infty + A e^{-\kappa L}/L^\alpha$, and fitting the various coefficients.  Figure~\ref{tp0scale} shows the resulting extrapolation for the density wave, parity, and string correlation functions.  For our extrapolation we use data with $L=32,64,128,256,512$, but for clarity do not plot the $L=512$ data. Notice the rapid convergence with system size, away from the phase boundaries.  We find equivalent results if we fix $L$, and fit $C_{ij}=O_\infty + A e^{-\kappa |i-j|}/|i-j|^\alpha$.

At non-zero $t^\prime/t$, the same procedure works for identifying the MI-HI transition.  For the other transitions, however, the results somewhat ambiguous, as the correlation length $\xi=1/\kappa$ becomes larger than the system size well before where $O_\infty$ vanishes, signaling a large critical region.  This behavior is consistent with the expected Kosterlitz-Thouless type physics, and we do not believe that we can accurately extrapolate to the thermodynamic limit.

\begin{figure}[t]
  \includegraphics[width=0.45\textwidth]{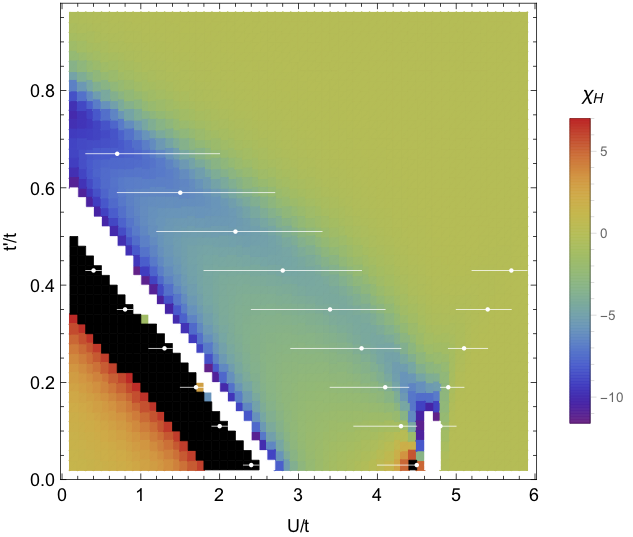}
    \includegraphics[width=0.45\textwidth]{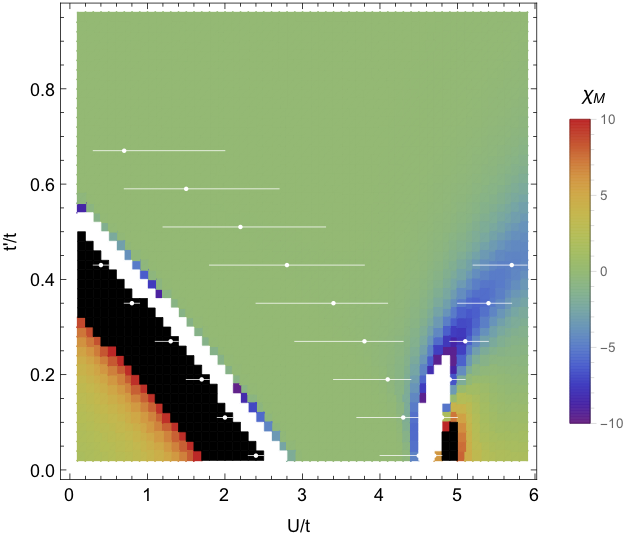}
  \caption{Curvatures, $\chi=t^2 (\partial^2O_L/\partial t^{\prime2} + \partial^2 O_L/\partial U^2)$ for $L=512$, where
  $O_L=C_{i=L/4,j=3L/4}$ is the correlation function at fixed distance. Left: string correlations, Right: parity correlations.
    Dots and error bars show phase boundaries determined from bipartite number fluctuations. White and black colors correspond to values outside the range of the scale bars.}
  \label{chi}
\end{figure}

We can, however, estimate the locations of the phase boundaries by looking for peaks in the curvature  $\chi=t^2 (\partial^2 O_L/\partial t^{\prime2} + \partial^2 O_L/\partial U^2)$.  Figure~\ref{chi} shows these susceptibilities for the string and parity correlation functions.  

\begin{figure}[t]
  \includegraphics[width=0.5\textwidth]{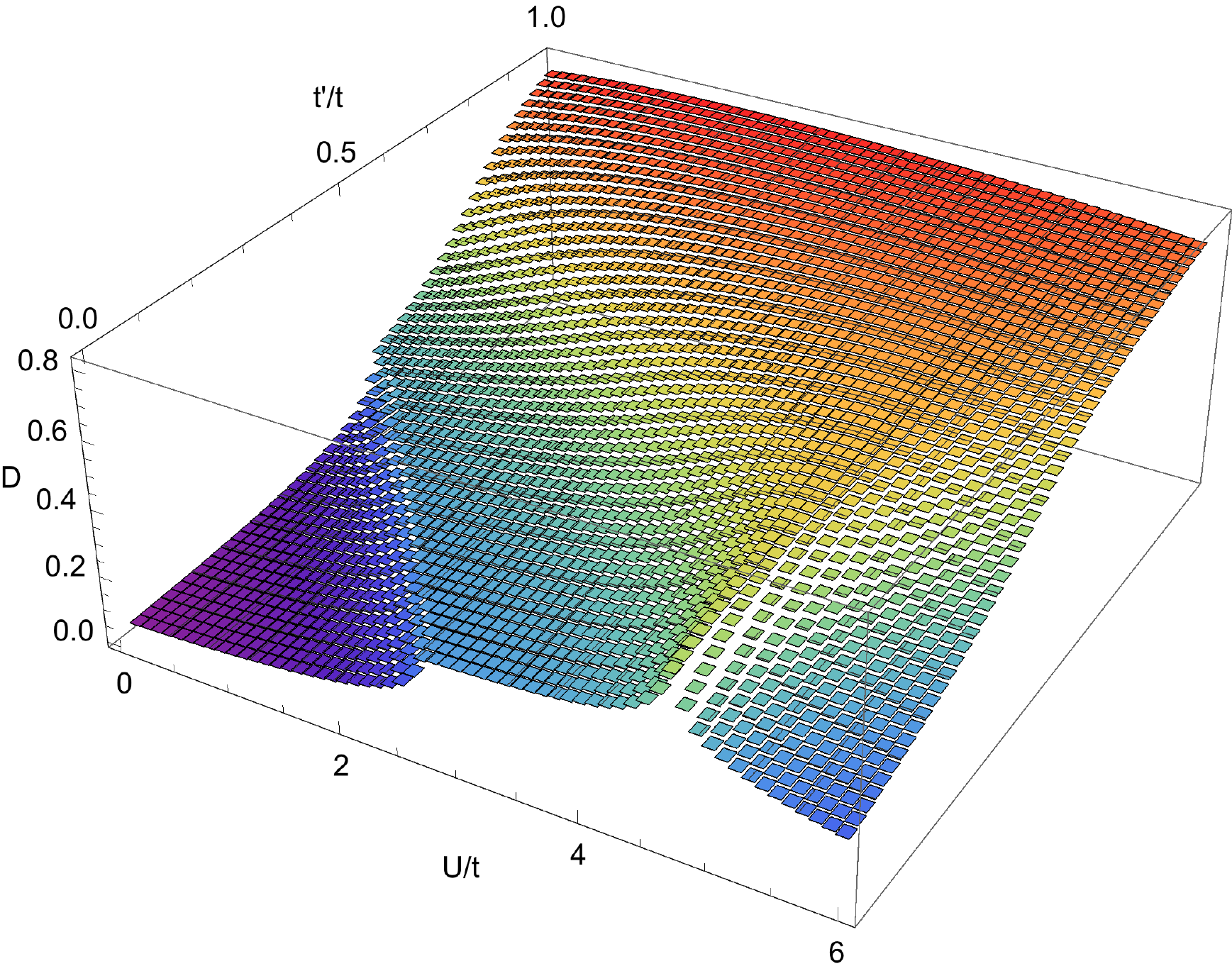}
  \caption{Bipartite number fluctuation $D_j =\langle N_{i<j}^2\rangle-\langle N_{i<j}\rangle^2$, where $N_{i<j}=\sum_{i=1}^{j-1} n_i$ for $j=L/2$ on a chain of length $L=256$.}
  \label{flucs}
\end{figure}

As discussed in the main text, we also estimate the locations of phase boundaries from the bipartite number fluctuations
$D_j =\langle N_{i<j}^2\rangle-\langle N_{i<j}\rangle^2$, where $N_{i<j}=\sum_{i=1}^{j-1} n_i$ is the number of particles to the left of site $j$.  In figure~\ref{flucs} we show $D_{L/2}$ for $L=256$.  Plateaus are indicative of bulk phases, while the regions of sharp variation represent our best estimates of the locations of the phase boundaries.  As we change system size, we find these transition regions become sharper, but we have not established if the width vanishes in the thermodynamic limit.  Nonetheless the width represents a conservative upper bound to our uncertainty about the location of the phase boundary.

\end{document}